\documentclass[useAMS,twocolumn,usenatbib,a4paper]{mn2e}

\usepackage{graphicx,enumitem, booktabs}

\usepackage{amssymb,amsmath, siunitx, wasysym}

\renewcommand{\vec}[1]{ \mathbf{#1} }

\begin{document}

\title[$f(R)$ gravity simulations of galaxy clusters]
{Scaling relations and mass bias in hydrodynamical $f(R)$ gravity simulations of galaxy clusters}
\author[Christian Arnold, Ewald Puchwein, Volker Springel]
{Christian Arnold$^{1}$, 
Ewald Puchwein$^{1,2}$,
Volker Springel$^{1,3}$
\vspace*{0.1cm}%
\\$^1$Heidelberger Institut f{\"u}r Theoretische Studien, Schloss-Wolfsbrunnenweg 35, 69118 Heidelberg, Germany
\\$^2$Institute of Astronomy and Kavli Institute for Cosmology, University of Cambridge, Madingley Road, Cambridge CB3 0HA, UK
\\$^3$Zentrum f\"ur Astronomie der Universit\"at Heidelberg, Astronomisches Recheninstitut, M\"{o}nchhofstr. 12-14, 69120 Heidelberg, Germany}
\date{\today}
\maketitle

\begin{abstract} 
We investigate the impact of chameleon-type $f(R)$ gravity models on
the properties of galaxy clusters and groups. Our $f(R)$ simulations
follow for the first time also the hydrodynamics of the intracluster
and intragroup medium. This allows us to assess how $f(R)$ gravity
alters the X-ray scaling relations of clusters and how hydrostatic and
dynamical mass estimates are biased when modifications of gravity are
ignored in their determination. We find that velocity dispersions and
ICM temperatures are both increased by up to $1/3$ in
$f(R)$ gravity in low-mass haloes, while the difference disappears in
massive objects. The mass scale of the transition depends on the
background value $f_{R0}$ of the scalar degree of freedom. These
changes in temperature and velocity dispersion alter the
mass-temperature and X-ray luminosity-temperature scaling relations
and bias dynamical and hydrostatic mass estimates that do not
explicitly account for modified gravity towards higher
values. Recently, a relative enhancement of X-ray compared to weak
lensing masses was found by the Planck
Collaboration~(\citeyear{planck2012}). We demonstrate that an
explanation for this offset may be provided by modified gravity and
the associated bias effects, which interestingly are of the required
size. Finally, we find that the abundance of subhaloes at fixed cluster
mass is only weakly affected by $f(R)$ gravity.
\end{abstract}


\section{Introduction}
\label{sec:introduction}

There are essentially two classes of models describing the late time
accelerated expansion of the Universe. In the first class, dubbed
``dark energy'' models, a new component is added to the
energy-momentum tensor of general relativity (GR). To drive the
observed accelerated expansion, this matter species or field has an
equation of state which provides a negative pressure. A prominent
example for this type of model is vacuum energy as considered in the
standard $\Lambda$ cold dark matter ($\Lambda$CDM) cosmology. In the second class, modifications to
general relativistic gravity are introduced that account for the
accelerated expansion. One of these so called ``modified gravity''
models is $f(R)$ gravity, in which a suitable scalar function $f(R)$
is added to the Ricci scalar $R$ in the gravitational part of the
action.

Since Einstein's general relativity is very well tested in the Solar
system, modified gravity models need a mechanism which ensures that
the modifications of gravity are suppressed locally so that
observational constraints are not violated. Several such models with
screening mechanisms have been constructed, such as the {\it
  Chameleon} \citep{khoury2004}, the {\it Symmetron}
\citep{hinterbichler2010}, the {\it Dilaton} \citep{gasperini2002},
and the {\it Vainshtein} \citep{vainshtein1972,deffayet2002}
mechanisms. In this work, we explore $f(R)$ gravity models which
exhibit a Chameleon-type screening \citep{husa2007}. The screening of
modified gravity effects is related to large non-linear perturbations
in the scalar degree of freedom which appear in this model. More
precisely, the screening length below which gravity is significantly
affected becomes negligibly small in these large perturbations which
are associated with matter over-densities like our Galaxy. Because of
these non-linearities, the effects on cosmic structure formation are
only accessible through detailed numerical simulations.

Recent numerical works on Chameleon-type $f(R)$ models of modified
gravity focused on quantities which can be studied with collisionless
simulations. The scale-dependent enhancement of the matter power
spectrum
\citep[e.g.][]{oyaizu2008,li2012,li2013,puchwein2013,llinares2013} and
halo mass function
\citep[e.g.][]{schmidt2009,ferraro2011,zhao2011b,lihu2011} was
analysed, as well as the impact of $f(R)$ gravity on cluster
concentrations \citep{lombriser2012b} and density profiles
\citep{lombriser2012}, halo velocity dispersions \citep{schmidt2010,
  lombriser2012b, lam2012}, redshift-space distortions
\citep{jennings2012}, and the integrated Sachs-Wolfe effect
\citep{cai2013}.

Our modified gravity simulation code, \textsc{mg-gadget}
(\citealt{puchwein2013}), allows us to follow baryonic physics and
modified gravity at the same time. This offers the opportunity to
investigate the ICM temperatures, the hydrostatic mass
bias, the X-ray luminosities and the thermal Sunyaev--Zeldovich (SZ) signals
of galaxy cluster and groups. Here, we assess how $f(R)$ gravity
affects these quantities, as well as cluster velocity dispersions,
subhalo abundances and dynamical mass estimates.

In Sect.~\ref{sec:fR_gravity}, we briefly summarize the main
properties of the $f(R)$ gravity model which we consider. An overview
of how our modified gravity simulation code works and what runs have
been performed with it is provided in Sect.~\ref{sec:simulations}. Our
results are presented in Sect.~\ref{sec:results}. We summarize our
findings and draw our conclusions in Sect.~\ref{sec:conclusions}.

\section{$\MakeLowercase{f}(R)$ gravity}
\label{sec:fR_gravity}

$f(R)$ gravity models generalize Einstein's general relativity by
adding a function $f(R)$ to the Ricci scalar $R$ in the gravitational
part of the action. The action is then given by
\begin{align}
 S=\int {\rm d}^4x \sqrt{-g} \left[ \frac{R+f(R)}{16\pi G} +\mathcal{L}_m \right],\label{action}
\end{align}
where $g$ is the determinant of the metric, $G$ is the gravitational
constant and $\mathcal{L}_m$ is the Lagrangian density of
matter. Demanding that the variation of this action with respect to
the metric vanishes leads to the modified Einstein equations
\citep{buchdahl1970}
\begin{align}
G_{\mu\nu} + f_R R_{\mu\nu}-\left( \frac{f}{2}-\Box f_R\right) g_{\mu\nu} - \nabla_\mu \nabla_\nu f_R = 8\pi G T_{\mu\nu} \label{Eequn},
\end{align}
where $G_{\mu\nu} = R_{\mu\nu} - R g_{\mu\nu}/2$ is the Einstein tensor and $f_R \equiv \text{d}f/\text{d}R$. Models which are compatible with observational constraints require $|f_R|\ll 1$. On scales much smaller than the horizon, the quasi-static approximation is valid \citep{oyaizu2008,noller2013} so that time derivatives can be neglected in the above equation. Together, this allows us to simplify the field equation for $f_R$ to (e.g. \citet{oyaizu2008}, also see Appendix \ref{app}) 
\begin{align}
 \nabla^2 f_R =  \frac{1}{3}\left(\delta R -8\pi G\delta\rho\right), \label{fRequn}
\end{align}
where $\delta R$ and $\delta \rho$ denote the perturbations in the
scalar curvature and matter density, respectively. Considering
Eq.~(\ref{Eequn}) in the Newtonian limit, a modified Poisson equation
for the gravitational potential is obtained (\citealt{husa2007}, also
see Appendix \ref{app})
\begin{align}
 \nabla^2 \Phi = \frac{16\pi G}{3}\delta\rho - \frac{1}{6} \delta R.\label{poisson}
\end{align}
In order to follow cosmic structure formation in $f(R)$ models, our code needs to solve the two partial differential equations (\ref{fRequn}) and (\ref{poisson}). The former equation is particularly challenging to solve due to its non-linearity. 

However let us first consider our choice of $f(R)$. Since GR is well
tested in the Solar system, modified gravity models should show the
same behaviour as GR in high density regions, or more precisely in our
local environment within the Milky Way. This is achieved in a class of
models which exhibit a chameleon mechanism, such as the model proposed
by \cite{husa2007},
\begin{align}
 f(R) = -m^2\frac{c_1\left(\frac{R}{m^2}\right)^n}{c_2\left(\frac{R}{m^2}\right)^n +1},
\end{align}
where $m^2 \equiv H_0^2\Omega_m$. For a suitable choice of the
parameters $c_1$, $c_2$ and $n$, the chameleon mechanism screens
$f(R)$ effects in high density regions. By also requiring
\begin{align}
 \frac{c_1}{c_2}=6\frac{\Omega_\Lambda}{\Omega_m} && \text{and}  &&  c_2\left(\frac{R}{m^2} \right)^n \gg 1, \label{eq:c1_c2_lambda}
\end{align}
an expansion history of the universe is obtained which closely mimics the one inferred with  a $\Lambda$CDM cosmological model \citep[see e.g.][]{husa2007}. In this scenario, the derivative of $f(R)$ is given by 
\begin{align}
 f_R=-n\frac{c_1\left(\frac{R}{m^2}\right)^{n-1}}{\left[c_2\left(\frac{R}{m^2}\right)^n+1\right]^2}\approx-n\frac{c_1}{c_2^2}\left(\frac{m^2}{R}\right)^{n+1},\label{fR}
\end{align}
where the second equality holds in the assumed limit
$c_2\left(\frac{R}{m^2} \right)^n \gg 1$. For a more convenient
characterization of a specific $f(R)$ model, the parameter set $c_1$
and $c_2$ can be replaced by the background value of $f_R$ at $z=0$,
$\bar{f}_{R0}$, as follows: the background curvature of a
Friedmann--Robertson--Walker universe is given by
\begin{align}
 \bar{R}=12 H^2 + 6\frac{\text{d}H}{\text{d}\ln a}H, 
\end{align}
which translates into
\begin{align}
  \bar{R}=3m^2\left[ a^{-3} + 4\frac{\Omega_\Lambda}{\Omega_m} \right]
\end{align}
for a flat $\Lambda$CDM expansion history. Plugging this equation for $a=1$ into Eq.~(\ref{fR}) and additionally demanding that the first equality in Eq.~(\ref{eq:c1_c2_lambda}) is satisfied constrains the parameters $c_1$ and $c_2$ completely for given values of $\Omega_\Lambda$, $\Omega_m$, $H_0$, $\bar{f}_{R0}$ and $n$. Hence, $\bar{f}_{R0}$ and $n$ can be used instead of $c_1$, $c_2$ and $n$ to completely specify the model. In the following sections, we will therefore describe the considered $f(R)$ models by their value of $\bar{f}_{R0}$. $n$ is fixed to $1$ in the simulations presented in this work. 

\section{The simulations}
\label{sec:simulations}

Our simulations were carried out with the modified gravity simulation
code \textsc{mg-gadget} (\citealt{puchwein2013}). The code is an
extension and modification of \textsc{p-gadget3}, which is itself
based on \textsc{gadget-2} (\citealt{springel2005c}). An advantage of
using \textsc{p-gadget3} as a basis for the modified gravity code is
that numerical models for a large number of physical processes, such
as hydrodynamics, gas cooling, star formation and associated feedback
processes are already implemented in this code. It is, hence, possible
to follow such baryonic processes and modified gravity at the same
time. Especially the possibility to account for hydrodynamics in
modified gravity simulations is essential for the analysis carried out
in this work.

Here, we provide only a very brief overview of how the
\textsc{mg-gadget} code solves the partial differential equations that
arise in $f(R)$ gravity. A detailed description of the code
functionality and the algorithms that are employed is given in
\citet{puchwein2013}. To solve the equation for $f_R$,
i.e. Eq.~(\ref{fRequn}), the code uses an iterative
multigrid-accelerated Newton-Gauss-Seidel relaxation scheme on an
adaptively refined mesh. This method is computationally efficient,
well suited for very non-linear equations and provides high spatial
resolution in high density regions, like in collapsed haloes. Note
however, that instead of solving directly for $f_R$, the code
iteratively computes $u \equiv \ln (f_R /\bar{f}_R(a))$. This ensures
that $f_R$ cannot attain unphysical positive values due to the finite
step size of the iterative solver, which makes the code numerically
more stable \citep[see also][]{oyaizu2008}.

Once $f(R)$ is known, the modified Poisson equation (\ref{poisson}) can be rewritten as
\begin{align}
\nabla^2 \Phi = 4\pi G (\delta\rho + \delta\rho_{\rm eff}),\label{rho}
\end{align}
where the effective mass density $\delta\rho_{\rm eff}$ encodes the modified gravity effects and is given by 
\begin{align}
\delta\rho_{\rm eff}= \frac{1}{3} \delta\rho - \frac{1}{24\pi G}\delta R.\label{rhoeff}
\end{align}
Adopting $n=1$, the following expression for $\delta R$ can be obtained from Eq.~(\ref{fR})
\begin{align}
\delta R = \bar{R}(a)\left( \sqrt{\frac{\bar{f}_R (a)}{f_R}}-1  \right).\label{deltaR}
\end{align} 
Hence, the code can compute the right-hand side of Eq.~(\ref{rho})
using the solution for $f(R)$, as well as the true mass density. The
resulting Poisson equation is subsequently solved with essentially the
same TreePM gravity solver which \textsc{p-gadget3} uses for standard
Newtonian gravity.  The hydrodynamics is followed with
\textsc{p-gadget3}'s entropy conserving smoothed particle
hydrodynamics scheme \citep{Springel2002}.

In the following, we will analyze four different sets of simulations,
each of them consisting of a $\Lambda$CDM and one or more $f(R)$
simulations which, all based on the same initial conditions. The
parameters of the simulations and the names by which we refer to them
are summarized in Table~\ref{tab:simulations}. Three of these sets
consist of pure dark matter, or more precisely collisionless
simulations, while the fourth set includes non-radiative
hydrodynamical runs as well.

\begin{table*}\centering
\begin{tabular}{lrrll} \toprule
Simulation & Box size $(\si{Mpc}/h)$ & Number of particles & Simulation type & Gravity\\ \midrule
DM-small & $\num{100}$ & $256^3$ & Collisionless & GR and $|\bar{f}_{R0}|=10^{-5}$\\
DM-large  & $\num{200}$ & $256^3$ & Collisionless & GR and $|\bar{f}_{R0}|=10^{-4/5/6}$\\
DM-high-res  & $\num{100}$ & $512^3$ & Collisionless & GR and $|\bar{f}_{R0}|=10^{-4}$\\
Nonrad  & $\num{200}$ & $2\times 256^3$ & Non-radiative hydro. & GR and $|\bar{f}_{R0}|=10^{-5}$\\
\bottomrule
\end{tabular}
\caption{Overview of the simulations we have performed. We will refer to the runs by the names provided in the first column. The box sizes are given in comoving coordinates.}
\label{tab:simulations}
\end{table*}

\section{Results}
\label{sec:results}

\subsection{Velocity dispersions of clusters and groups}

\begin{figure*}
\centerline{\includegraphics[width=\linewidth]{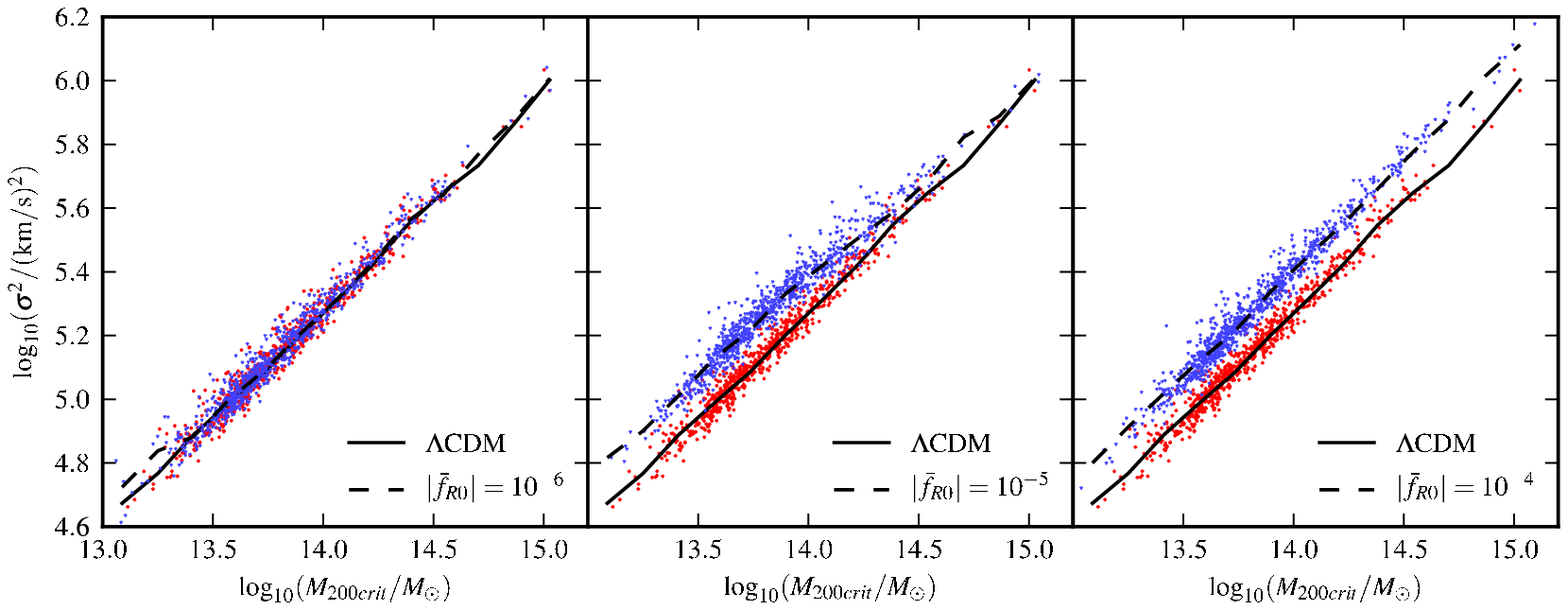}}
\caption{One-dimensional velocity dispersion of the `DM-large'
  simulation particles as a function of halo mass $M_{\rm 200crit}$
  for $\Lambda$CDM (\textit{red dots, solid line}) and $f(R)$
  (\textit{blue dots, dashed line}) cosmologies and different
  $\bar{f}_{R0}$ values. Lines show the medians of the binned data.}
\label{fig:veldisp}
\end{figure*}

As a first step in exploring the dynamical properties of our simulated
clusters and groups, we have computed the one-dimensional velocity
dispersions of the `DM-large' simulation particles within $r_{\rm
  200crit}$, which is the radius of a sphere that is centred on the
potential minimum of the cluster and inside which the mean density is
200 times the critical density of the Universe. To this end, the haloes
have been identified and their potential minima have been found with
the {\sc subfind} code \citep{Springel2001}. The results, which are
based on the DM-large simulations, are displayed in
Fig.~\ref{fig:veldisp} for GR, as well as for$|\bar{f}_{R0}|=10^{-6}$,
$10^{-5}$ and $10^{-4}$.

The figure illustrates both the enhancement of the velocity dispersion
due to modified gravity effects, as well as their screening in massive
haloes. For $|\bar{f}_{R0}|=10^{-4}$ the velocity dispersion is
increased by about $1/3$ with respect to $\Lambda$CDM over the whole
mass range. In this model, even the gravitational potential wells of
clusters are not deep enough for the chameleon mechanism to become
fully effective. This increment is, hence, theoretically expected. In
particular, combining equations (\ref{rho}) and (\ref{rhoeff}) for
$\delta R\approx 0$, i.e. in the low-curvature regime, leads to forces
larger by a factor of $4/3$ with respect to GR, which translates into
an increase of the squared velocity dispersions by roughly the same
factor.

For $|\bar{f}_{R0}|=10^{-5}$, the mass threshold for the onset of the
chameleon mechanism is smaller. This is reflected by our results. At
low masses, which correspond to more shallow potential wells, the
velocity dispersion is again increased by a factor of $\sim 4/3$
compared to $\Lambda$CDM. At about $10^{14} M_\odot$ the chameleon
mechanism sets in and the difference between $f(R)$ and $\Lambda$CDM
decreases until it vanishes almost completely roughly above $10^{14.5}
M_\odot$.

In the $|\bar{f}_{R0}|=10^{-6}$ cosmology, the $f(R)$ effects on
gravity are screened essentially over the whole mass range shown in the
figure. Thus, there is almost no difference in the median curves of
the $\Lambda$CDM and $f(R)$ runs. A small deviation is, however,
present at the low mass end. This presumably indicates the
transition to the unscreened low-curvature regime. Overall, these
results are in good agreement with the findings of \cite{schmidt2010}.

\begin{figure}
\centerline{\includegraphics[width=\linewidth]{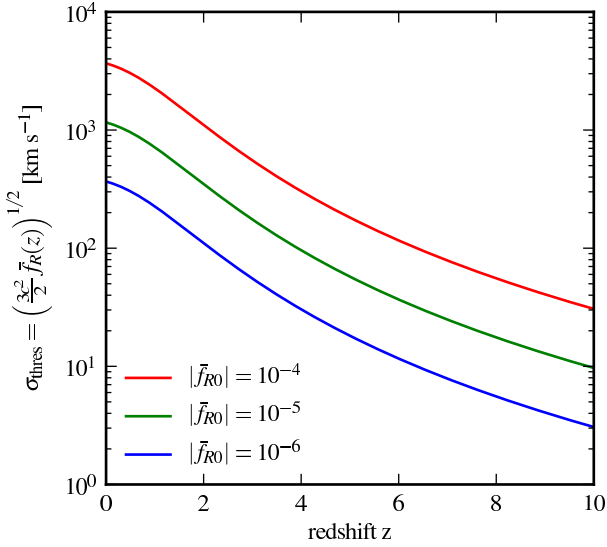}}
\caption{Approximate theoretical threshold value in the
  three-dimensional halo velocity dispersion for the screening of
  $f(R)$ effects on gravity due to the chameleon mechanism. In haloes
  exceeding this threshold, GR gravity is expected to be restored.}
\label{fig:screening}
\end{figure}

One can obtain a simple analytic estimate of the velocity dispersion
threshold above which the chameleon mechanism screens modified gravity
effects. Note that in the low curvature regime $\nabla^2 f_R \approx
-\frac{8 \pi G}{3c^2} \delta \rho \approx -\frac{2}{3c^2} \nabla^2
\phi_{\rm N}$, where $\phi_{\rm N}$ is the Newtonian gravitational
potential. From this one finds $\delta f_R \approx -\frac{2}{3c^2}
\phi_{\rm N}$ in the unscreened regime. The chameleon effect becomes
effective once strong non-linearities appear. According to
Eq.~(\ref{deltaR}), this happens when $|\delta f_R|$ approaches
$|\bar{f}_R|$. Hence, chameleon screening is active for $|\phi_{\rm
  N}| \gtrsim \frac{3c^2}{2} |\bar{f}_R|$
\citep[e.g.][]{husa2007,cabre2012}. For the sake of simplicity we
ignore for the moment factors of approximately $\sim \sqrt{(4/3)}$ due
to modified gravity effects when translating the Newtonian potential
to a three-dimensional halo velocity dispersion $\sigma$. Assuming
$\sigma_{\rm 3D} \approx \sqrt{\phi_{\rm N}}$, this results in a
threshold value for the onset of chameleon screening of $\sigma_{\rm
  3D} \gtrsim \sqrt{\frac{3c^2}{2} \bar{f}_R}$. Figure
\ref{fig:screening} displays this quantity as a function of redshift for
models with $|\bar{f}_{R0}|=10^{-6}$, $10^{-5}$ and $10^{-4}$.

For $|\bar{f}_{R0}|=10^{-4}$ and $z=0$, the onset of screening is
expected at $\sigma_{1\text{D}}=\sigma_{3\text{D}}/\sqrt{3} \approx
\sqrt{10^{6.6}} \, {\rm km\, s^{-1}}$, which is even larger than the values found
in our most massive simulated galaxy clusters. The theoretical value
for $|\bar{f}_{R0}|=10^{-5}$ at $z=0$,
i.e. ${\sigma_{1\text{D}}}=\sqrt{10^{5.6}}\, {\rm km\, s^{-1}}$, is in
good agreement with the position of the transition region in the
simulation. The theoretical value for the onset of screening in the
$|\bar{f}_{R0}|=10^{-6}$ cosmology is $\sigma_{1\text{D}} \approx
\sqrt{10^{4.6}} \, {\rm km\, s^{-1}}$, which is compatible with the slight
increase in the velocity dispersion that we find for low mass objects
in the corresponding simulation.

Note, however, that the simple derivation presented above neglects the effects of environment. In particular, the Newtonian gravitational potential is not only affected by an object's mass but also by its surroundings. Thus, even objects with masses below the derived screening threshold can be screened, if they reside in a high density region. This effect could result in increased scatter of the properties of low mass objects in $f(R)$ gravity. Massive galaxy clusters are in contrast not expected to be strongly affected.

\subsection{Temperatures of the intracluster and intragroup medium}

\begin{figure}
\centerline{\includegraphics[width=\linewidth]{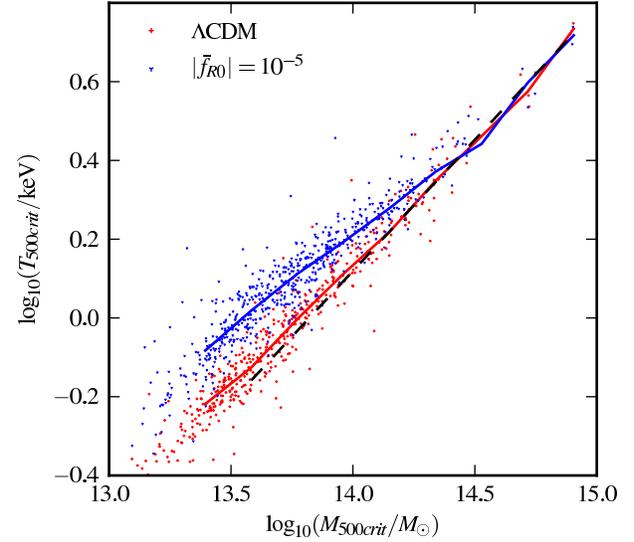}}
\caption{Relation between mass-weighted temperatures and group/cluster
  masses within $r_{\rm 500crit}$ in the `Nonrad' simulation in a
  $\Lambda$CDM and a $|\bar{f}_{R0}|=10^{-5}$ cosmology. The
  \textit{solid} lines show the median of the binned data, the
  \textit{dashed} line indicates the slope of the self-similar scaling
  relation $T\propto M^{2/3}$.}
\label{fig:temp}
\end{figure}

As the temperature of the intracluster or intragroup medium is closely
related to the halo velocity dispersion, we expect to find a similar
behaviour in the mass-temperature scaling relation. For the `Nonrad'
simulation, this is indeed the case as illustrated in Figure
\ref{fig:temp}, which displays the relation between the group/cluster
masses and mass-weighted temperatures, $T_{\rm MW}=\sum T_{\rm particle}
m_{\rm particle}/ \sum m_{\rm particle}$ within a radius that encloses a mean
density of $500$ times the critical density of the Universe. As
theoretically expected, the non-radiative $\Lambda$CDM relation
follows the slope of the self-similar prediction \citep{kaiser1986},
i.e. $T\propto M^{2/3}$, which is indicated by the \textit{dashed}
line in the figure.
 
In $f(R)$ gravity, the $M$-$T$ relation deviates from the $\Lambda$CDM
result. Like the velocity dispersions, the temperatures are boosted by
about $30\% - 40\%$ with respect to the standard cosmology at masses
below approximately $10^{13.8} M_\odot$. This increment is again
comparable to the enhancement of the gravitational forces. At about
$10^{13.8} M_\odot$ screening as implied by the chameleon mechanism
sets in. This reduces the difference in the $\Lambda$CDM and $f(R)$
relations with increasing mass until the curves coincide for $M_{\rm
  500crit} \gtrsim 10^{14.5} M_\odot$.

\begin{figure}
\centerline{\includegraphics[width=\linewidth]{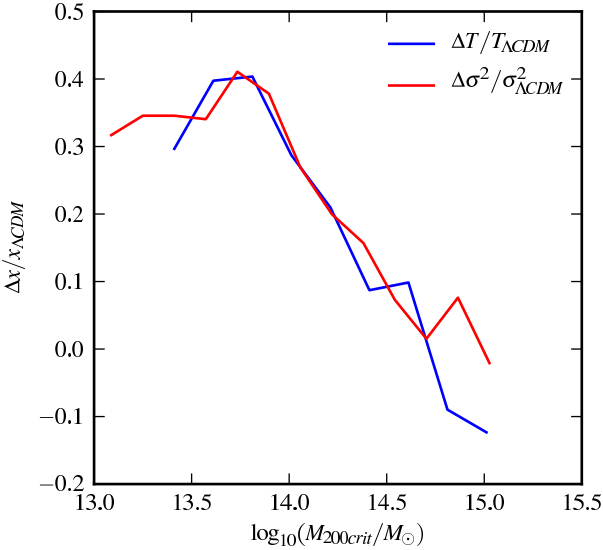}}
\caption{Relative difference between $\Lambda$CDM and
  $|\bar{f}_{R0}|=10^{-5}$ cosmologies in median velocity dispersion
  and mass-weighted temperature. All quantities were computed within
  $r_{\rm 200crit}$. The velocity dispersion was calculated for all
  friends-of-friends groups while the temperature was computed only
  for those consisting of at least $2000$ particles.}
\label{fig:compveldisptemp}
\end{figure}

The masses shown in Figures \ref{fig:veldisp} and \ref{fig:temp} were
calculated for different spherical overdensity thresholds, i.e. within
different radii. To be able to directly compare the enhancement of
mass-weighted temperatures and halo velocity dispersions in $f(R)$
gravity, Figure \ref{fig:compveldisptemp} shows the relative
difference in the median curves for both quantities. Here, all values
were calculated within $r_{\rm 200crit}$. The figure visualizes the
theoretically expected effects on both of these quantities. The curves
coincide almost perfectly.

\subsection{Mass bias}

\subsubsection{Dynamical mass estimates}

A standard method for inferring the masses of distant galaxy clusters
and groups is to relate the line-of-sight (LOS) velocity dispersion of
their member galaxies to their mass by applying the virial
theorem. Here, we investigate how these masses are biased in $f(R)$
gravity if modified gravity effects are not specifically corrected for
in the analysis, i.e. when the standard relation between velocity
dispersion and halo mass is used. To this end, we calculate the LOS
velocity dispersion of the subhaloes identified by \textsc{subfind} for
massive haloes in our $\Lambda$CDM and $f(R)$ simulations. Based on
them, we then estimate the dynamical masses of the haloes with the
method described in \cite{bahcall1981}. As the velocity dispersions
are enhanced in $f(R)$ gravity one expects the dynamical masses to be
higher too. To assess the bias in these mass estimates, we compare the
dynamical masses to the true masses, which are simply calculated by
summing up the masses of the simulation particles within
$r_{\rm 200crit}$.

\begin{figure*}
\centerline{\includegraphics[width=\linewidth]{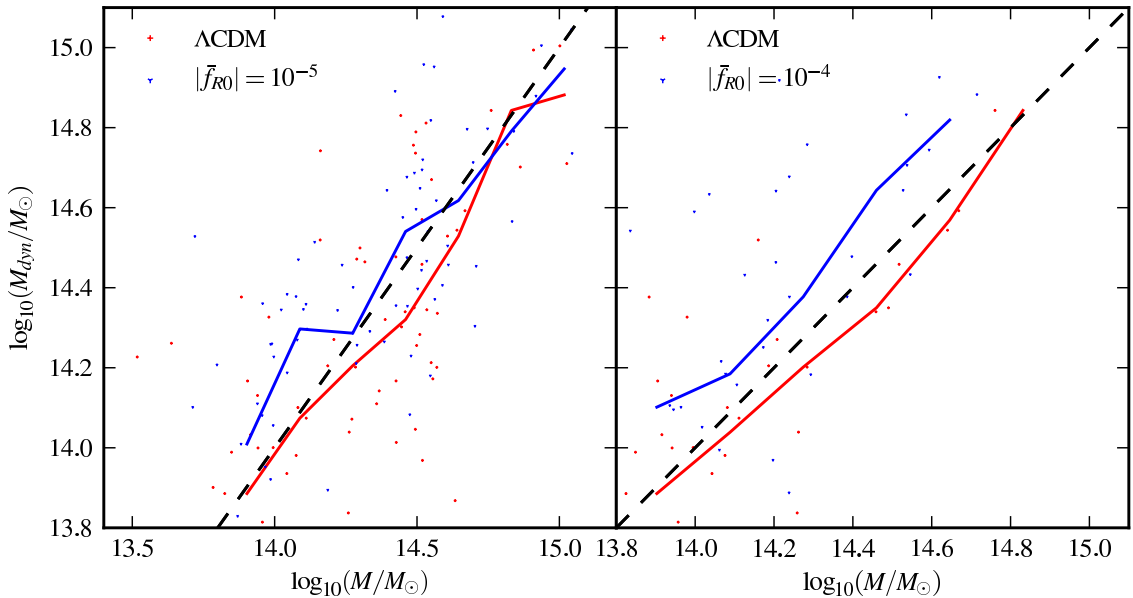}}
\caption{Dynamical vs. real mass for $\Lambda$CDM and $f(R)$
  cosmology. The left-hand panel shows the $40$ largest groups of the
  `DM-small' and `DM-large' simulations, each for $\Lambda$CDM and
  $|\bar{f}_{R0}|=10^{-5}$ cosmologies. Both simulations use $256^3$
  particles but differ in box size ($\SI{100}{Mpc}/h$ and
  $\SI{200}{Mpc}/h$, respectively). The right-hand panel displays the $40$
  largest haloes of the `DM-high-res' simulation carried out in
  $\Lambda$CDM and $|\bar{f}_{R0}|=10^{-4}$ cosmologies. All results
  are based on collisionless simulations. The dynamical masses are
  computed with the method described in \protect\cite{bahcall1981}
  using the velocity dispersion of the subhaloes identified by
  \textsc{subfind} to represent the velocity dispersion of cluster
  galaxies. \textit{Solid} lines indicate the median of the binned
  data, \textit{dashed} lines the 1:1 relation.}
\label{fig:dynmass}
\end{figure*}

Figure \ref{fig:dynmass} shows the dynamical mass -- true mass
relation in $\Lambda$CDM and for $|\bar{f}_{R0}|=10^{-5}$ and
$10^{-4}$. For $|\bar{f}_{R0}|=10^{-5}$, we combine results form the
`DM-small' and the `DM-large' simulations to cover a larger range in
halo mass. The true and dynamical masses of the 40 largest groups of
each simulation are indicated by dots in the plot, \textit{solid}
lines show the medians for each cosmology. As expected, the dynamical
masses exhibit a similar behaviour as the velocity dispersion. At low
masses the dynamical mass estimates in an $f(R)$ cosmology are too
high due to the larger dispersion of the subhalo velocities. This
clearly demonstrates that in order to obtain accurate masses based on
the velocity dispersions, one has to modify the virial theorem instead
of using the standard relation which is valid only in GR. At higher
masses the chameleon effect sets in and the $f(R)$ dynamical masses
are compatible with the GR results. However, even the $\Lambda$CDM
curve does not accurately recover the real mass (\textit{dashed line})
in an intermediate mass region, i.e. around $M_{\rm 200crit} \approx
10^{14.5} M_\odot$. This is likely caused by the large scatter implied
by the low number of objects.

For $|\bar{f}_{R0}|=10^{-4}$, shown in the figure's right-hand panel, the
behaviour at low masses is the same as in the left hand plot but the
dynamical masses are overestimated in the whole mass range displayed in
the figure. This is due to the much deeper potential wells that are
required for the onset of the chameleon mechanism for larger
$|\bar{f}_{R0}|$ values. In the considered mass range they are simply
not deep enough for the screening of $f(R)$ effects to become
effective. These results are consistent with the behaviour of the
velocity dispersions for different $|\bar{f}_{R0}|$ presented in
\cite{schmidt2010}.

\begin{figure}
\centerline{\includegraphics[width=\linewidth]{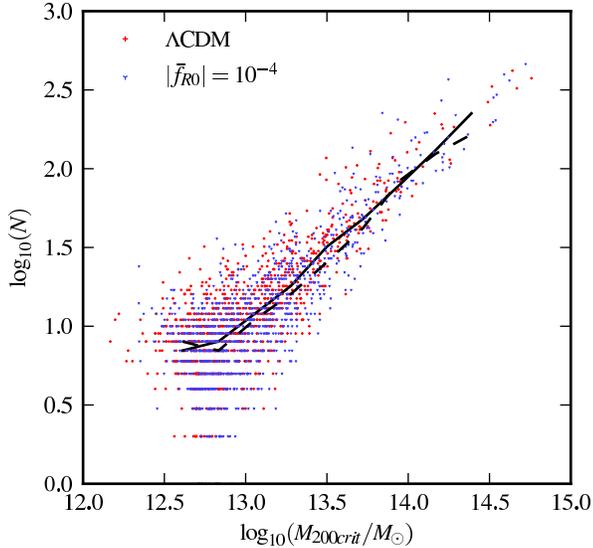}}
\caption{Number $N$ of resolved subhaloes as a function of the mass of
  the parent halo. Only objects for which the main halo is resolved by
  at least 2000 simulation particles in the `DM-high-res' simulations
  are included. The lines show the medians of the binned data for
  $\Lambda$CDM (\textit{solid line}) and $f(R)$ (\textit{dashed
    line}).}
\label{fig:richness}
\end{figure}

The increment in dynamical masses is caused by the higher subhalo
velocity-dispersion in $f(R)$ gravity. In contrast, there is only a
small difference in subhalo abundance between modified gravity and GR
in our simulations. This is illustrated in Figure~\ref{fig:richness}.

\subsubsection{Hydrostatic masses}
\label{sec:hydro_mass}

\begin{figure}
\centerline{\includegraphics[width=\linewidth]{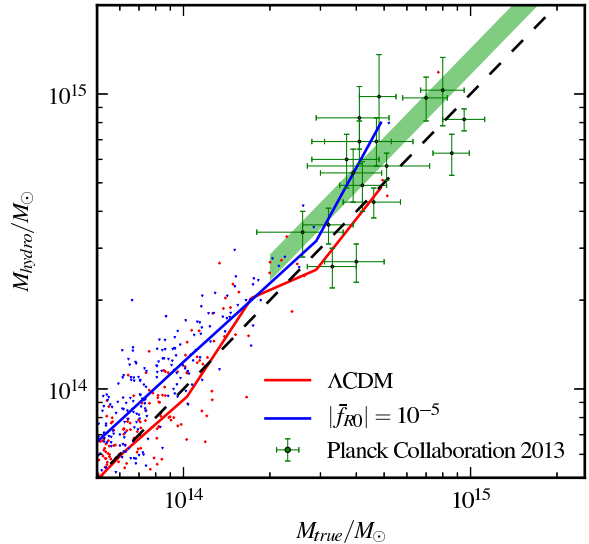}}
\caption{Relation between hydrostatic mass estimates and the true mass
  for relaxed (see text) clusters in the $\Lambda$CDM (\textit{red})
  and $|\bar{f}_{R0}|=10^{-5}$ (\textit{blue}) simulations,
  respectively. \textit{Solid} lines show the median of the binned
  data, the \textit{dashed} line indicates the 1:1 relation. The
  \textit{green} symbols with error bars are the data points of the
  Planck Collaboration~(\citeyear{planck2012}) (hydrostatic and weak
  lensing masses) and the \textit{green shaded} region represent their
  best fit region. All values are measured within $r_{\rm
    500crit}$. For the $f(R)$ model, mass estimates were computed
  assuming an observer is unaware of modified gravity effects.}
\label{fig:hydromass}
\end{figure}

Our hydrodynamical simulations offer the opportunity to investigate
cluster mass estimates that are based on the properties of the
ICM, a method which is used as a standard
technique to interpret X-ray observations of galaxy clusters. These so
called hydrostatic masses are compared to the true mass in Figure
\ref{fig:hydromass}. The medians are indicated by \textit{solid}
lines. The \textit{dashed} line shows the 1:1 relation. Hydrostatic
masses are estimated from the pressure gradient at $r_{\rm 500crit}$.

To obtain results which are as realistic as possible for the given
simulation, both thermal and non-thermal pressure contributions were
considered. The thermal pressure is calculated from the temperatures
of the gas particles while the non-thermal part is computed from the
velocities of the simulation particles relative to their host
halo. The latter corresponds to bulk motions in the ICM. Effectively, the pressure is, thus, computed based on the sum
of the thermal and the kinetic energy (in the object's rest frame) of
the ICM.

The non-thermal pressure defined in this way can, however, be easily
overestimated in merging clusters. Furthermore, in those objects the
hydrostatic equilibrium assumption is likely violated. To prevent our
results from being strongly affected by mergers, we introduce a
criterion for the selection of clusters. Only objects whose average
kinetic particle energy around $r_{\rm 500crit}$ does not exceed $0.5$
times the thermal energy of the particles are considered in the
analysis. Additionally, criteria for identifying relaxed systems,
based on considering the centre-of-mass displacement and the mass in
substructures, were applied. These criteria are similar to those used
in \cite{neto2007}. If the distance from the centre-of-mass to the
minimum of the gravitational potential is larger than $0.105 \times
r_{\rm 200crit}$, the object is considered as not relaxed and excluded
from the analysis. The same is the case if the mass of the
substructures found by \textsc{subfind} exceeds $10\%$ of the total
cluster mass. As the plot shows, these criteria ensure that
hydrostatic and true mass are in good agreement in the $\Lambda$CDM
model. It was also checked that the selection criteria do not
bias the results in $f(R)$ gravity. In particular, we found that there
is no significant difference in the fraction of relaxed objects at fixed 
halo mass in $f(R)$ gravity compared to $\Lambda$CDM. The ratio of 
thermal to non-thermal pressure remains also unchanged.

For the $f(R)$ cosmology, the hydrostatic masses were computed in the
same way as for $\Lambda$CDM. In particular, modified gravity effects
were not explicitly accounted for in the mass estimates. These
estimates, thus, correspond to hydrostatic masses computed from
observations assuming that the observer is unaware of the presence of
modifications of gravity. As expected, this results in an overestimate
of the true mass due to the modified relation between the mass
distribution and the gravitational potential. As a cautionary remark,
we would like to add that this mass bias should not be confused with
hydrostatic mass biases that may arise in $\Lambda$CDM due to
violations of the hydrostatic equilibrium condition or due to
unaccounted non-thermal pressure components.

To compare the simulations to recent observational data, the results
found by the Planck Collaboration (\citeyear{planck2012}) were added
to the plot (\textit{green symbols}). Like for the simulations, the
hydrostatic masses $M^{\rm hydro}_{500}$ are shown on the figure's
vertical axis, while the horizontal axis displays weak lensing masses
$M^{\rm WL}_{500}$. Except for observational errors in the weak
lensing analysis, the latter can be considered to represent the true
mass as lensing deflection angles and mass estimates are not altered
by $f(R)$-gravity effects in models with $|\bar{f}_{R0}| \ll 1$ \citep[see Appendix \ref{app}
  and][]{zhao2011}. The best fit region from that work, $M^{\rm
  WL}_{500}=(0.78 \pm 0.08) M^{\rm hydro}_{500}$, is shaded in green.

Like in our $f(R)$ simulations, the Planck
Collaboration~(\citeyear{planck2012}) found hydrostatic masses which
are larger than the corresponding weak lensing masses. If this result
is, indeed, substantiated by future studies and not caused by some
observational bias, modified gravity could provide a theoretical
explanation for it. One should keep in mind, however, that there are
observational uncertainties that might also cause such a bias
(some of them are discussed in \citealt{planck2012}, as well as in \citealt{applegate2012}). Furthermore, there are also authors that find that hydrostatic masses are smaller than weak lensing masses \citep{mahdavi2013}.
Finally, we have to acknowledge that  $|\bar{f}_{R0}|=10^{-5}$ might already be in tension with Solar system constraints \citep{husa2007}. However, given that the onset of screening is not visible in Fig.~\ref{fig:hydromass} even for the most massive simulated clusters (see discussion below), it is quite possible that higher hydrostatic cluster masses also appear for somewhat lower $|\bar{f}_{R0}|$.

Comparing the mass difference in Figure \ref{fig:hydromass} to the
previous plots, one might be surprised why the effects of a chameleon
screening of $f(R)$ gravity for objects with masses above $10^{14}
M_\odot$ are not clearly visible, as this was the case for the
previously analysed quantities and $|\bar{f}_{R0}|=10^{-5}$. The
reason is most likely that the quantities in the other plots are
calculated by averaging over the whole volume within the considered
radius, while the hydrostatic masses are computed using the pressure
and potential gradients at a relatively large specific radius, i.e. at
$r_{\rm 500crit}$. It is, hence, more sensitive to the cluster outskirts,
where the potential is not as deep as in the central region. This
shifts the transition of the screened regime to larger cluster
masses. As a consequence, the lack of more massive clusters in our
simulations prevents us from seeing this transition in
Fig.~\ref{fig:hydromass}.

The increased dynamical and hydrostatic masses in $f(R)$ gravity are
consistent with the theoretical expectations. Since gravity is
enhanced by a factor of up to $4/3$ with respect to GR one expects an
enhancement of mass estimates of the same order
(\citealt{schmidt2009}). Figures \ref{fig:dynmass} and
\ref{fig:hydromass}, indeed, show that an increment of this order is
present in our results.

\subsection{Scaling relations}

\subsubsection{The $L$-$T$ scaling relation}


Our `Nonrad' simulations follow the evolution of the density and
temperature distributions of the ICM. This allows us to investigate
the X-ray properties of our simulated clusters. In particular, we
calculate emission-weighted temperatures and X-ray luminosities for
all clusters and groups found by \textsc{subfind}. To calculate the
luminosity, metal line emission was neglected as the exact shape of
the cooling function is unlikely to have a qualitative impact on the
comparison between different cosmological models. Furthermore, the
simulations do not follow the metal enrichment of the ICM, so that ad
hoc assumptions about the metallicity would be necessary to include
line emission. Finally, to be able to directly compare our simulated
scaling relation to observations, it would be necessary to account for
additional baryonic physics that is important in this context, like
star formation and feedback from active galactic nuclei \citep[see
  e.g.][]{puchwein2008}.

We, hence, stick to the simple assumption of free--free bremsstrahlung
and compute the luminosities according to $L\propto\sum \rho m
\sqrt{T}$, where $\rho$ is the gas density, $T$ the gas temperature
and $m$ the gas particle mass. The sum extends over all gas particles
within $r_{\rm 500crit}$. The emission-weighted temperatures are
estimated by weighting the temperatures of the gas particles within
the same radius with their estimated X-ray luminosity.
  
\begin{figure}
\centerline{\includegraphics[width=\linewidth]{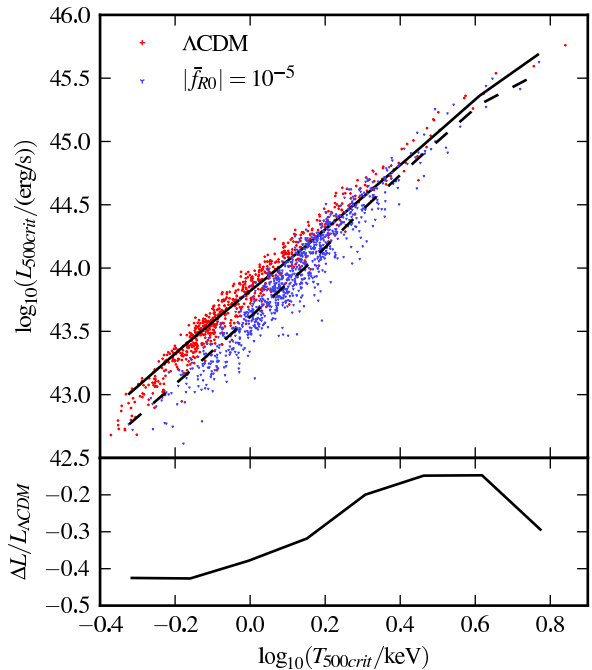}}
\caption{\textit{Upper panel:} X-ray luminosities of clusters and
  groups as a function of their emission-weighted temperature. Results
  are shown for a $|\bar{f}_{R0}|=10^{-5}$ (\textit{blue, dashed
    line}) and a $\Lambda$CDM (\textit{red, solid line})
  cosmology. The values are computed within $r_{\rm 500crit}$ for all
  objects resolved by at least 2000 simulation particles. The lines
  show the medians of the binned data. \textit{Lower panel:} Relative
  difference of the medians shown in the upper panel.}
\label{fig:luminosity}
\end{figure}

Figure \ref{fig:luminosity} shows the relation between X-ray
luminosity and temperature for all groups and clusters which are
resolved by at least 200 particles. Results are shown both for a
$\Lambda$CDM and a $|\bar{f}_{R0}|=10^{-5}$ cosmology. The
luminosities at fixed temperature are lower in $f(R)$ gravity compared
to $\Lambda$CDM. This is not unexpected: as shown in Figure
\ref{fig:temp}, the temperature of the ICM at a given halo mass is
larger in $f(R)$ gravity. Or in other words, the mass of an object at
given temperature will be lower in $f(R)$. Given that the luminosity
is roughly proportional to $\sim M \sqrt{T}$, a lower mass $M$
translates to a lower X-ray luminosity. Also note that the slope of
the cooling rate as a function of temperature would be even lower in
the relevant range if metal cooling were accounted for. Including
metal line cooling would, thus, not qualitatively change our
results. As expected, the difference between the models decreases at
high temperatures where the chameleon mechanism becomes effective.

\subsubsection{The $Y_{\rm SZ}$-$M$ scaling relation}


Another observational probe of the intracluster and intragroup medium
is the spectral distortion of the cosmic microwave background that is
caused by foreground galaxy clusters and groups. This distortion,
known as the thermal SZ effect, can be described
by the Compton-$y$ parameter, which is a scaled projection of the gas
pressure in the intervening cluster or group. Integrating the
Compton-$y$ parameter over the projected extent of the objects on the
sky yields $Y_{\rm SZ}$, which correlates well with halo mass.

\begin{figure}
\centerline{\includegraphics[width=\linewidth]{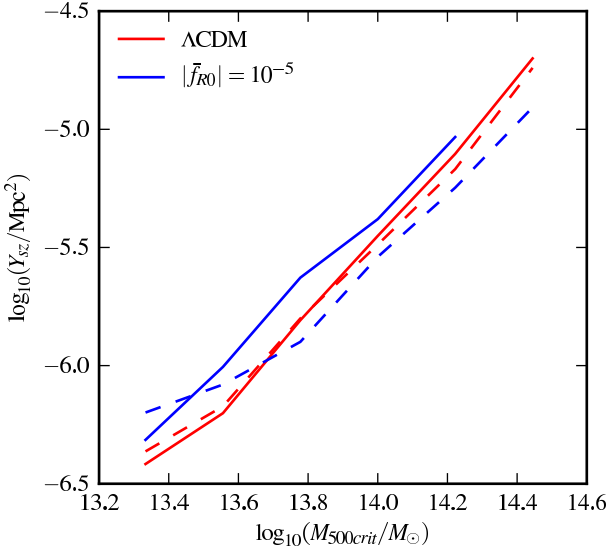}}
\caption{Integrated thermal SZ signal $Y_{\rm SZ}$ as a
  function of true (\textit{solid lines}) and hydrostatic
  (\textit{dashed lines}) mass for the $\Lambda$CDM (\textit{red}) and
  the $|\bar{f}_{R0}|=10^{-5}$ (\textit{blue}) cosmologies. The lines
  show the median values for all groups which fit the criteria for
  relaxed systems (see Sec.~\ref{sec:hydro_mass}). Hydrostatic masses
  were estimated in the same way as for Figure \ref{fig:hydromass}.}
\label{fig:sz}
\end{figure}

The relation between $Y_{\rm SZ}$ and mass is shown in
Figure~\ref{fig:sz}. It is computed using the temperatures and masses
of the gas particles in the `Nonrad' simulation outputs of both the
$\Lambda$CDM and the $|\bar{f}_{R0}|=10^{-5}$ runs. The integrated
Compton parameter is plotted against true and hydrostatic masses,
where the latter have been computed in the same way as for
Figure~\ref{fig:hydromass}. Comparing the $Y_{\rm SZ}$-true mass
relations, we find larger $Y_{\rm SZ}$ values in $f(R)$ than in
$\Lambda$CDM. This is expected since $Y_{\rm SZ}$ depends on the
electron temperatures which are larger in $f(R)$ gravity. Like for
many of the previously analysed quantities, the difference between the
models decreases at about ${10^{14}}{M_\odot}$ due to the chameleon
mechanism becoming effective there.

As expected from Fig.~\ref{fig:hydromass}, the $Y_{\rm
  SZ}$-hydrostatic mass relation, which could be probed by a
comparison of X-ray and SZ data, is basically identical to the $Y_{\rm
  SZ}$-true mass relation in $\Lambda$CDM. In contrast, the $Y_{\rm
  SZ}$-mass relation strongly depends on the mass measure in $f(R)$
gravity. Larger hydrostatic masses in the $f(R)$ model shift the curve
to the right. Interestingly, the $Y_{\rm SZ}$-hydrostatic mass
relation is rather similar in $\Lambda$CDM and $f(R)$. Larger
hydrostatic masses almost compensate the stronger SZ
signal. This results in a shift mostly along the relation, rather than
perpendicular to it. In observed scaling relations, it will therefore
be easier to see effects of $f(R)$ gravity in the $Y_{\rm SZ}$-lensing
mass relation.
 
\section{Summary and conclusions}
\label{sec:conclusions}

We have performed the first analysis of galaxy clusters and groups in
cosmological hydrodynamical simulations of $f(R)$ gravity models,
using the \citet{husa2007} parametrization. In addition, we have
studied collisionless runs of the same model, as well as reference
$\Lambda$CDM simulations. The hydrodynamical simulations allowed us to
explore the effects of modified gravity on the ICM and
its observable properties, as well as on hydrostatic mass estimates
and X-ray and SZ scaling relations. The dynamics of cluster galaxies,
as traced by self-bound subhaloes, were investigated both in
hydrodynamical and collisionless runs. The effects of modified gravity
on dynamical mass estimates were determined. Our main findings are the following.

\begin{itemize}
\setlength{\itemindent}{0.15cm}

\item The dark matter velocity dispersions of low-mass haloes are
  boosted by roughly a factor $\sqrt{4/3}$ in $f(R)$ gravity compared
  to $\Lambda$CDM. This is consistent both with previous works
  (\citealt{schmidt2010}) and theoretical expectations, given that
  gravitational forces are increased by a factor of $4/3$ in these
  objects.

\item The maximum halo mass for which this enhanced velocity
  dispersion is observed depends of the value $|\bar{f}_{R0}|$, as it
  controls in which objects the chameleon mechanism is effective. For
  $|\bar{f}_{R0}|=10^{-6}$ modified gravity effects are screened in
  almost the whole mass range we have explored, so that no changes in
  the velocity dispersions compared to $\Lambda$CDM are observed. For
  $|\bar{f}_{R0}|=10^{-5}$, velocity dispersions are increased below
  $\sim 10^{14.5} M_\odot$ and show little difference above this
  value. For $|\bar{f}_{R0}|=10^{-4}$, there is no screening even in
  massive clusters, so that velocity dispersions are boosted in all
  haloes in this case. We show that this trend is consistent with
  theoretical expectations.

\item The mass-temperature scaling relation is affected by $f(R)$
  gravity. More precisely, in unscreened, i.e. less massive haloes the
  temperature at fixed mass increases with respect to $\Lambda$CDM. In
  particular, we find that the gas temperatures and the variance of
  the dark matter velocities are boosted by the same halo
  mass-dependent factor.

\item $f(R)$ gravity not only increases the velocity dispersion of
  dark matter particles, but also the dispersion of subhalo
  velocities. This translates into a bias in dynamical mass
  estimates. More precisely, halo masses estimated based on subhalo
  velocities overpredict the true mass in unscreened haloes unless
  modified gravity effects are explicitly accounted for in the
  analysis. In massive haloes in which the chameleon mechanism is
  effective, the dynamical masses are in good agreement with the true
  masses. The abundance of subhaloes is only very mildly affected by
  $f(R)$ gravity.

\item Hydrostatic masses are also increased in $f(R)$ gravity compared
  to $\Lambda$CDM if modified gravity effects are not explicitly
  accounted for in the mass estimate. In contrast to the previously
  analysed quantities, we do not see a transition from the unscreened
  to the screened regime in the hydrostatic mass-true mass relation in
  our non-radiative hydrodynamical simulation. This has most likely
  the following reason: hydrostatic masses were computed from the
  pressure gradient at a specific, relatively large radius, namely
  $r_{\rm 500crit}$. There the gravitational potential is not as deep as
  in the central cluster or group region. Thus, at these large radii
  the transition between the unscreened and the screened regime is
  shifted towards more massive objects which are not present in the
  analysed simulations.

\item While hydrostatic masses are biased in $f(R)$ gravity, lensing
  mass estimates are not affected. This results in an offset in the
  X-ray mass-lensing mass relation of roughly the same magnitude as
  recently found by the Planck
  Collaboration~(\citeyear{planck2012}). If their finding is
  substantiated by future studies, modified gravity could provide an
  explanation for this offset.

\item We find a lower normalization of the X-ray
  luminosity-temperature scaling relation in $f(R)$ gravity. The
  increment in X-ray luminosities in $f(R)$ gravity is overcompensated
  by the boosted temperatures. This results in a lower normalization
  of the $L-T$ relation for objects in which modified gravity is not
  efficiently screened.

\item The SZ signal of our simulated clusters and
  groups is affected by $f(R)$ gravity as well. The effect on the
  $Y_{\rm SZ}-M$ relation depends, however, significantly on the mass
  measure which is used. When plotted against hydrostatic masses, the
  relation in $f(R)$ gravity is only shifted along the GR-curve since
  both quantities are enhanced. The $f(R)$ effects on the SZ signal
  are much more clearly visible if $Y_{\rm SZ}$ is displayed as a
  function of true or lensing mass.

\end{itemize}

Overall, our analysis demonstrates that $f(R)$ gravity significantly
affects the velocity dispersions, virial temperatures and scaling
relations of unscreened haloes. Furthermore, an observer who is unaware
of modifications of gravity would obtain biased mass estimates both
from a dynamical analysis, as well as based on the assumption of
hydrostatic equilibrium in a Newtonian potential.

In the future, it will be interesting to include further baryonic
physics in cosmological hydrodynamical simulations of modified
gravity, as well as to push to higher resolution. This will allow
investigating modified gravity effects on galaxies and galaxy
populations self-consistently, thereby complementing work based on
semi-analytical galaxy formation models \citep[see][]{fontanot2013}.

\section*{Acknowledgements}

We are grateful to Marco Baldi for helpful discussions. E.P. and
V.S. acknowledge support by the DFG through Transregio
33, `The Dark Universe'. 
E.P. also acknowledges support by the ERC grant `The Emergence of 
Structure during the epoch of Reionization'.

\appendix
\section{The modified Poisson equation and lensing masses in $\MakeLowercase{f}(R)$ gravity}\label{app}

The aim of this appendix is to derive the modified Poisson equation
for $f(R)$ gravity (Eq.~\ref{poisson}) and the equation for the
scalar degree of freedom (Eq.~\ref{fRequn}), as well as to show that
the relations between mass density and lensing deflection angles are
identical in $f(R)$ gravity and GR. This ensures that lensing masses
are not biased by $f(R)$ modifications of gravity.

Adopting the Newtonian gauge and assuming for simplicity a spatially
flat background, the line element can be written as
\begin{align}
 {\rm d}s^2 = a(\eta)^2 \left[-(1+2\Phi) {\rm d} \eta^2 + (1-2\Psi) {\rm d}\vec{x}^2 \right],\label{linee}
\end{align}
where $\eta$ denotes conformal time and is related to cosmic time $t$
by $a(\eta) {\rm d}\eta \equiv {\rm d}t$. In this metric gravitational
lensing is governed by $\Phi_L=(\Phi+\Psi)/2$
(\citealt{bartelmann2010, zhao2011}). To simplify the calculations, we
define $F=f_R+1$ and rewrite Eq.~(\ref{Eequn}) as
\begin{align}
 F R_{\mu\nu} - \frac{1}{2} f g_{\mu\nu}- \frac{1}{2} R g_{\mu\nu} - \nabla_\mu \nabla_\nu F +g_{\mu\nu}\Box F = 8\pi G T_{\mu\nu}\label{FRequn},
\end{align}
where the d'Alembert operator is defined by $\Box =
g^{\mu\nu}\nabla_\mu \nabla_\nu$. Taking the trace of
Eq.~(\ref{FRequn}) leads to
\begin{align}
 F R -2 f - 2 R + 3 \Box F = 8\pi G T.\label{FRtrace}
\end{align}

\subsection{The equation for $f_R$}

As $|f_R| \ll 1$ in all models we consider, we can approximate
$F\approx 1$ in the first term of Eq.~(\ref{FRtrace}). Subtracting the
background equation from the result and neglecting a $\delta f$ term,
as $|\delta f| \approx |f_R\delta R| \ll |\delta R|$, yields
\begin{align}
 \Box f_R =  \frac{1}{3}\left(\delta R -8\pi G\delta\rho\right),\label{dalambertfR}
\end{align}
where the energy-momentum tensor was assumed to be that of a
pressure-less fluid, so that $\delta T=-\delta \rho$, and $\delta\rho$
is the perturbation in physical density. In the quasi-static limit,
i.e. neglecting all time derivatives and assuming instantaneous
propagation of gravity \citep[see][for a discussion of the validity of
  this assumption]{noller2013}, this equation turns into
\begin{align}
 \frac{1}{a^2}\nabla^2 f_R = \nabla^2_{\rm phys} f_R = \frac{1}{3}\left(\delta R -8\pi G\delta\rho\right), \label{laplacefR}
\end{align}
where $\nabla^2$ and $\nabla^2_{\rm phys}$ denote the Laplace operators with respect to comoving and physical coordinates, respectively. The second equality is identical to Eq.~(\ref{fRequn}), except that we have omitted the subscript in $\nabla^2_{\rm phys}$ in the latter equation for the sake of brevity.

\subsection{The modified Poisson equation for the gravitational potential}

Taking the $00$-component of Eq.~(\ref{FRequn}) and plugging it into the metric Eq.~(\ref{linee}) yields
\begin{align}
 F R_{00} +\frac{1}{2}(f + R) a^2 (1 + 2\Phi) - \frac{1 + 2\Phi}{1-2\Psi}\nabla^2 F = 8\pi G \rho \, a^2.
\end{align}
The second time derivative of $F$, corresponding to the fourth term of
Eq.~(\ref{FRequn}), has here been canceled by the time component of the
D'Alembert operator. The energy-momentum tensor was assumed to
have the pressure-less perfect fluid form, i.e. $T_{00} = \rho \,
a^2$.

In the following we adopt the weak field limit, i.e. we assume $|\Phi|\ll1$ and $|\Psi|\ll1$, and consider models with $|f_R| \ll 1$, so that we can approximate $F\approx 1$ in the first term. After subtracting the background equation, we find
\begin{align}
 \delta R_{00} + \frac{1}{2} \delta R \, a^2 - \nabla^2 f_R = 8\pi G \delta \rho \, a^2,
\end{align}
where  $|\delta f| \approx |f_R\delta R| \ll |\delta R|$ has been used. Plugging in (\ref{laplacefR}), one obtains
\begin{align}
 \delta R_{00}= \frac{16 \pi G}{3} \delta\rho \, a^2 -\frac{1}{6} \delta R \, a^2. \label{R00}
\end{align}

A similar calculation for the space-space components of (\ref{FRequn}) leads to
\begin{align}
 \delta R_{ii} -\frac{1}{2}\delta R \, a^2 - \nabla_i \nabla_i \delta f_R + \nabla^2 \delta f_R = 0,
\end{align}
where we have again assumed $|\Phi|\ll1$, $|\Psi|\ll1$ and have adopted the quasi-static approximation. Summing over the spatial components, one obtains
\begin{align}
 \sum_{i=1}^3 \delta R_{ii} -\frac{3}{2} \delta R \, a^2 + 2 \nabla^2 \delta f_R =0,
\end{align}
Using (\ref{laplacefR}) and sorting terms yields
\begin{align}
 \sum_{i=1}^3 \delta R_{ii} =  \frac{16 \pi G}{3}\delta\rho \, a^2 + \frac{5}{6} \delta R \, a^2. \label{Rs}
\end{align}

To obtain the Ricci tensor for the considered metric (\ref{linee}), the Christoffel symbols 
\begin{align}
 \Gamma^\kappa_{\mu\nu}=\frac{1}{2} g^{\kappa\alpha}
 (g_{\alpha\nu,\mu}+g_{\mu\alpha,\nu}-g_{\mu\nu,\alpha})\label{christoffel}
\end{align}
must be calculated.
Denoting derivatives with respect to conformal time and spatial comoving coordinates as $x' \equiv \partial_\eta x$ and $x_{,j} \equiv \partial_j x$, the components can be written as
\begin{align}
\Gamma^0_{00}& =\frac{a'}{a}+\Phi' , \quad & \Gamma^i_{00}&=\Phi_{,i}
\;\; , \nonumber\\ \Gamma^0_{0i}& =\Phi_{,i} \;\; , \quad &
\Gamma^i_{0j}&=\left(\frac{a'}{a}-\Psi'\right)\delta^i_j ,\nonumber
\end{align}
\vspace{-0.5cm}
\begin{align}
\Gamma^0_{ij}& =\left(\frac{a'}{a} \left[1-2(\Phi+\Psi)\right] -\Psi'
\right)\delta_{ij}\nonumber  , \\ \Gamma^i_{jk}&
=\Psi_{,i}\delta_{jk}-\Psi_{,j}\delta^i_k-\Psi_{,k}\delta^i_j .
\end{align}

The Ricci tensor can be computed from the connection forms as
\begin{align}
 R_{\mu\nu} = \Gamma^\kappa_{\mu\nu,\kappa}-\Gamma^\kappa_{\kappa\mu,\nu}+\Gamma^\kappa_{\kappa\gamma}\Gamma^\gamma_{\nu\mu}-\Gamma^\kappa_{\nu\gamma}\Gamma^\gamma_{\kappa\mu}.
\end{align}
Neglecting second and higher order terms in $\Phi$ and $\Psi$ the diagonal elements turn out to be
\begin{align}
 &R_{00} &= &-3\mathcal{H}' +3 \Psi'' + \nabla^2\Phi +3\mathcal{H}(\Phi'+\Psi'),\\
&R_{ii} &= &(\mathcal{H}'+2\mathcal{H}^2)\nonumber\\& &+ &\left[ -\Psi''+\nabla^2\Psi-\mathcal{H}(\Phi'+5\Psi')-(2\mathcal{H}'+4\mathcal{H}^2)(\Phi+\Psi) \right]\nonumber\\& &+ &\nabla_i\nabla_i(\Psi-\Phi),
\end{align}
in the coordinates defined by Eq.~(\ref{linee}). Here, $\mathcal{H}=a'/a$ is the conformal Hubble function. In the quasi-static regime and on scales much smaller than the Hubble radius, we can neglect all time derivatives of $\Phi$ and $\Psi$, as well as factors of $\mathcal{H}$ and its derivatives. This yields
\begin{align}
 &\delta R_{00} &= & \,\,\, \nabla^2 \Phi,\label{R002}\\
 &\sum_{i=1}^3 \delta R_{ii} &= & \,\,\, 3 \, \nabla^2 \Psi + \nabla^2(\Psi-\Phi) = 4 \, \nabla^2 \Psi - \nabla^2 \Phi\label{Rs2}.
\end{align} 

Combining equations (\ref{R00}) and (\ref{R002}) results in the modified Poisson equation (Eq.~\ref{poisson}) for the gravitational potential
\begin{align}
 \frac{1}{a^2}\nabla^2\Phi = \nabla^2_{\rm phys} \Phi = \frac{16 \pi G}{3}\delta\rho-\frac{1}{6}\delta R\label{phi}.
\end{align}

\subsection{Gravitational lensing in $f(R)$ gravity}

Using this result together with  (\ref{Rs}) and (\ref{Rs2}) leads to a similar relation for $\Psi$
\begin{align}
 \frac{1}{a^2}\nabla^2\Psi = \nabla^2_{\rm phys}\Psi = \frac{8 \pi G}{3}\delta\rho+\frac{1}{6}\delta R\label{psi}.
\end{align}
Thus the potential $\Phi_L=(\Psi+\Phi)/2$, which governs gravitational light deflection, satisfies 
\begin{align}
 \nabla^2_{\rm phys}\Phi_L=\frac{\nabla^2_{\rm phys}\Phi+\nabla^2_{\rm phys}\Psi}{2} = 4\pi G \delta\rho = \nabla^2_{\rm phys} \phi_{\rm N},
\end{align}
i.e. it satisfies the same standard Poisson equation as the Newtonian
gravitational potential $\phi_{\rm N}$, and is hence unchanged despite
the $f(R)$ effects on $\Phi$ and $\Psi$. As a consequence, the
gravitational lensing deflection angle
\begin{align}
 \boldsymbol{\alpha}=\frac{2}{c^2} \int \boldsymbol{\nabla}_\perp \left( \frac{\Phi+\Psi}{2}\right)\, {\rm d}l = 
\frac{2}{c^2} \int \boldsymbol{\nabla}_\perp \phi_{\rm N}\, {\rm d}l,
\end{align}
is the same as in GR. Here $\boldsymbol{\nabla}_\perp$ is the
component of the gradient with respect to physical coordinates which
is perpendicular to the line of sight. The line element ${\rm d}l$
corresponds to proper distance. Weak lensing mass estimates are, thus,
not affected by $f(R)$ gravity in 
models with $|\bar{f}_{R0}| \ll 1$.

\bibliographystyle{mn2efixed}
\bibliography{paper}

\begin{thebibliography}{37}
\expandafter\ifx\csname natexlab\endcsname\relax\def\natexlab#1{#1}\fi

\bibitem[{{Applegate} {et~al}\mbox{.}(2012){Applegate}, {von der Linden},
  {Kelly}, {Allen}, {Allen}, {Burchat}, {Burke}, {Ebeling}, {Mantz}, \&
  {Morris}}]{applegate2012}
{Applegate} D.~E. {et~al.}, 2012, ArXiv e-prints: 1208.0605

\bibitem[{{Bahcall} \& {Tremaine}(1981)}]{bahcall1981}
{Bahcall} J.~N., {Tremaine} S., 1981, \apj, 244, 805

\bibitem[{{Bartelmann}(2010)}]{bartelmann2010}
{Bartelmann} M., 2010, Classical and Quantum Gravity, 27, 233001

\bibitem[{{Buchdahl}(1970)}]{buchdahl1970}
{Buchdahl} H.~A., 1970, \mnras, 150, 1

\bibitem[{{Cabr{\'e}} {et~al}\mbox{.}(2012){Cabr{\'e}}, {Vikram}, {Zhao},
  {Jain}, \& {Koyama}}]{cabre2012}
{Cabr{\'e}} A., {Vikram} V., {Zhao} G.-B., {Jain} B., {Koyama} K., 2012, \jcap,
  7, 34

\bibitem[{{Cai} {et~al}\mbox{.}(2013){Cai}, {Li}, {Cole}, {Frenk}, \&
  {Neyrinck}}]{cai2013}
{Cai} Y.-C., {Li} B., {Cole} S., {Frenk} C.~S., {Neyrinck} M., 2013, ArXiv
  e-prints: 1310.6986

\bibitem[{{Deffayet} {et~al}\mbox{.}(2002){Deffayet}, {Dvali}, {Gabadadze}, \&
  {Vainshtein}}]{deffayet2002}
{Deffayet} C., {Dvali} G., {Gabadadze} G., {Vainshtein} A., 2002, \prd, 65,
  044026

\bibitem[{{Ferraro} {et~al}\mbox{.}(2011){Ferraro}, {Schmidt}, \&
  {Hu}}]{ferraro2011}
{Ferraro} S., {Schmidt} F., {Hu} W., 2011, \prd, 83, 063503

\bibitem[{{Fontanot} {et~al}\mbox{.}(2013){Fontanot}, {Puchwein}, {Springel},
  \& {Bianchi}}]{fontanot2013}
{Fontanot} F., {Puchwein} E., {Springel} V., {Bianchi} D., 2013, \mnras, 436,
  2672

\bibitem[{{Gasperini} {et~al}\mbox{.}(2002){Gasperini}, {Piazza}, \&
  {Veneziano}}]{gasperini2002}
{Gasperini} M., {Piazza} F., {Veneziano} G., 2002, \prd, 65, 023508

\bibitem[{{Hinterbichler} \& {Khoury}(2010)}]{hinterbichler2010}
{Hinterbichler} K., {Khoury} J., 2010, Physical Review Letters, 104, 231301

\bibitem[{{Hu} \& {Sawicki}(2007)}]{husa2007}
{Hu} W., {Sawicki} I., 2007, \prd, 76, 064004

\bibitem[{{Jennings} {et~al}\mbox{.}(2012){Jennings}, {Baugh}, {Li}, {Zhao}, \&
  {Koyama}}]{jennings2012}
{Jennings} E., {Baugh} C.~M., {Li} B., {Zhao} G.-B., {Koyama} K., 2012, \mnras,
  425, 2128

\bibitem[{{Kaiser}(1986)}]{kaiser1986}
{Kaiser} N., 1986, \mnras, 222, 323

\bibitem[{{Khoury} \& {Weltman}(2004)}]{khoury2004}
{Khoury} J., {Weltman} A., 2004, \prd, 69, 044026

\bibitem[{{Lam} {et~al}\mbox{.}(2012){Lam}, {Nishimichi}, {Schmidt}, \&
  {Takada}}]{lam2012}
{Lam} T.~Y., {Nishimichi} T., {Schmidt} F., {Takada} M., 2012, Physical Review
  Letters, 109, 051301

\bibitem[{{Li} {et~al}\mbox{.}(2013){Li}, {Hellwing}, {Koyama}, {Zhao},
  {Jennings}, \& {Baugh}}]{li2013}
{Li} B., {Hellwing} W.~A., {Koyama} K., {Zhao} G.-B., {Jennings} E., {Baugh}
  C.~M., 2013, \mnras, 428, 743

\bibitem[{{Li} {et~al}\mbox{.}(2012){Li}, {Zhao}, {Teyssier}, \&
  {Koyama}}]{li2012}
{Li} B., {Zhao} G.-B., {Teyssier} R., {Koyama} K., 2012, \jcap, 1, 51

\bibitem[{{Li} \& {Hu}(2011)}]{lihu2011}
{Li} Y., {Hu} W., 2011, \prd, 84, 084033

\bibitem[{{Llinares} {et~al}\mbox{.}(2013){Llinares}, {Mota}, \&
  {Winther}}]{llinares2013}
{Llinares} C., {Mota} D.~F., {Winther} H.~A., 2013, arXiv: 1307.6748

\bibitem[{{Lombriser} {et~al}\mbox{.}(2012{\natexlab{a}}){Lombriser}, {Koyama},
  {Zhao}, \& {Li}}]{lombriser2012b}
{Lombriser} L., {Koyama} K., {Zhao} G.-B., {Li} B., 2012{\natexlab{a}}, \prd,
  85, 124054

\bibitem[{{Lombriser} {et~al}\mbox{.}(2012{\natexlab{b}}){Lombriser},
  {Schmidt}, {Baldauf}, {Mandelbaum}, {Seljak}, \& {Smith}}]{lombriser2012}
{Lombriser} L., {Schmidt} F., {Baldauf} T., {Mandelbaum} R., {Seljak} U.,
  {Smith} R.~E., 2012{\natexlab{b}}, \prd, 85, 102001

\bibitem[{{Mahdavi} {et~al}\mbox{.}(2013){Mahdavi}, {Hoekstra}, {Babul},
  {Bildfell}, {Jeltema}, \& {Henry}}]{mahdavi2013}
{Mahdavi} A., {Hoekstra} H., {Babul} A., {Bildfell} C., {Jeltema} T., {Henry}
  J.~P., 2013, \apj, 767, 116

\bibitem[{{Neto} {et~al}\mbox{.}(2007){Neto}, {Gao}, {Bett}, {Cole}, {Navarro},
  {Frenk}, {White}, {Springel}, \& {Jenkins}}]{neto2007}
{Neto} A.~F. {et~al.}, 2007, \mnras, 381, 1450

\bibitem[{Noller {et~al}\mbox{.}(2014)Noller, von Braun-Bates, \&
  Ferreira}]{noller2013}
Noller J., von Braun-Bates F., Ferreira P.~G., 2014, Phys. Rev. D, 89, 023521

\bibitem[{{Oyaizu}(2008)}]{oyaizu2008}
{Oyaizu} H., 2008, \prd, 78, 123523

\bibitem[{{Planck Collaboration} {et~al}\mbox{.}(2013){Planck Collaboration},
  {Ade}, {Aghanim}, {Arnaud}, {Ashdown}, {Atrio-Barandela}, {Aumont},
  {Baccigalupi}, {Balbi}, {Banday}, {Barreiro}, {Bartlett}, {Battaner},
  {Battye}, {Benabed}, {Bernard}, {Bersanelli}, {Bhatia}, {Bikmaev},
  {B{\"o}hringer}, {Bonaldi}, {Bond}, {Borgani}, {Borrill}, {Bouchet},
  {Bourdin}, {Brown}, {Bucher}, {Burenin}, {Burigana}, {Butler}, {Cabella},
  {Cardoso}, {Carvalho}, {Chamballu}, {Chiang}, {Chon}, {Clements},
  {Colafrancesco}, {Coulais}, {Cuttaia}, {Da Silva}, {Dahle}, {Davis}, {de
  Bernardis}, {de Gasperis}, {Delabrouille}, {D{\'e}mocl{\`e}s}, {D{\'e}sert},
  {Diego}, {Dolag}, {Dole}, {Donzelli}, {Dor{\'e}}, {Douspis}, {Dupac},
  {Efstathiou}, {En{\ss}lin}, {Eriksen}, {Finelli}, {Flores-Cacho}, {Forni},
  {Frailis}, {Franceschi}, {Frommert}, {Galeotta}, {Ganga},
  {G{\'e}nova-Santos}, {Giard}, {Giraud-H{\'e}raud}, {Gonz{\'a}lez-Nuevo},
  {G{\'o}rski}, {Gregorio}, {Gruppuso}, {Hansen}, {Harrison},
  {Hern{\'a}ndez-Monteagudo}, {Herranz}, {Hildebrandt}, {Hivon}, {Hobson},
  {Holmes}, {Huffenberger}, {Hurier}, {Jagemann}, {Juvela}, {Keih{\"a}nen},
  {Khamitov}, {Kneissl}, {Knoche}, {Kunz}, {Kurki-Suonio}, {Lagache},
  {Lamarre}, {Lasenby}, {Lawrence}, {Le Jeune}, {Leach}, {Leonardi}, {Liddle},
  {Lilje}, {Linden-V{\o}rnle}, {L{\'o}pez-Caniego}, {Luzzi},
  {Mac{\'{\i}}as-P{\'e}rez}, {Maino}, {Mandolesi}, {Maris}, {Marleau},
  {Marshall}, {Mart{\'{\i}}nez-Gonz{\'a}lez}, {Masi}, {Matarrese}, {Matthai},
  {Mazzotta}, {Meinhold}, {Melchiorri}, {Melin}, {Mendes}, {Mitra},
  {Miville-Desch{\^e}nes}, {Montier}, {Morgante}, {Munshi}, {Natoli},
  {N{\o}rgaard-Nielsen}, {Noviello}, {Osborne}, {Pajot}, {Paoletti},
  {Partridge}, {Pearson}, {Perdereau}, {Perrotta}, {Piacentini}, {Piat},
  {Pierpaoli}, {Piffaretti}, {Platania}, {Pointecouteau}, {Polenta},
  {Ponthieu}, {Popa}, {Poutanen}, {Pratt}, {Prunet}, {Puget}, {Rachen},
  {Rebolo}, {Reinecke}, {Remazeilles}, {Renault}, {Ricciardi}, {Ristorcelli},
  {Rocha}, {Rosset}, {Rossetti}, {Rubi{\~n}o-Mart{\'{\i}}n}, {Rusholme},
  {Sandri}, {Savini}, {Scott}, {Starck}, {Stivoli}, {Stolyarov}, {Sudiwala},
  {Sunyaev}, {Sutton}, {Suur-Uski}, {Sygnet}, {Tauber}, {Terenzi},
  {Toffolatti}, {Tomasi}, {Tristram}, {Valenziano}, {Van Tent}, {Vielva},
  {Villa}, {Vittorio}, {Wandelt}, {Weller}, {White}, {Yvon}, {Zacchei}, \&
  {Zonca}}]{planck2012}
{Planck Collaboration} {et~al.}, 2013, \aap, 550, A129

\bibitem[{{Puchwein} {et~al}\mbox{.}(2013){Puchwein}, {Baldi}, \&
  {Springel}}]{puchwein2013}
{Puchwein} E., {Baldi} M., {Springel} V., 2013, \mnras, 436, 348

\bibitem[{{Puchwein} {et~al}\mbox{.}(2008){Puchwein}, {Sijacki}, \&
  {Springel}}]{puchwein2008}
{Puchwein} E., {Sijacki} D., {Springel} V., 2008, \apjl, 687, L53

\bibitem[{{Schmidt}(2010)}]{schmidt2010}
{Schmidt} F., 2010, \prd, 81, 103002

\bibitem[{{Schmidt} {et~al}\mbox{.}(2009){Schmidt}, {Lima}, {Oyaizu}, \&
  {Hu}}]{schmidt2009}
{Schmidt} F., {Lima} M., {Oyaizu} H., {Hu} W., 2009, \prd, 79, 083518

\bibitem[{{Springel}(2005)}]{springel2005c}
{Springel} V., 2005, \mnras, 364, 1105

\bibitem[{{Springel} \& {Hernquist}(2002)}]{Springel2002}
{Springel} V., {Hernquist} L., 2002, \mnras, 333, 649

\bibitem[{{Springel} {et~al}\mbox{.}(2001){Springel}, {White}, {Tormen}, \&
  {Kauffmann}}]{Springel2001}
{Springel} V., {White} S.~D.~M., {Tormen} G., {Kauffmann} G., 2001, \mnras,
  328, 726

\bibitem[{Vainshtein(1972)}]{vainshtein1972}
Vainshtein A., 1972, Physics Letters B, 39, 393

\bibitem[{{Zhao} {et~al}\mbox{.}(2011{\natexlab{a}}){Zhao}, {Li}, \&
  {Koyama}}]{zhao2011b}
{Zhao} G.-B., {Li} B., {Koyama} K., 2011{\natexlab{a}}, \prd, 83, 044007

\bibitem[{{Zhao} {et~al}\mbox{.}(2011{\natexlab{b}}){Zhao}, {Li}, \&
  {Koyama}}]{zhao2011}
{Zhao} G.-B., {Li} B., {Koyama} K., 2011{\natexlab{b}}, Physical Review
  Letters, 107, 071303

\end{thebibliography}
\end{document}